\documentclass[fdp,a4paper,fleqn%
]{w-art}
\usepackage{times,cite,w-thm}
\theoremstyle{plain}

\theoremstyle{definition}

\usepackage[]{graphicx}

\usepackage{amsmath,amssymb}
\newcommand{\rmd}{\mathrm{d}}
\newcommand{\Ps}{P}
\newcommand{\Qs}{Q}
\newcommand{\Heff}{{H_\mathrm{eff}}}

\begin{document}
\DOIsuffix{theDOIsuffix}
\Volume{55}
\Month{03}
\Year{2012}
\pagespan{1}{}
\Receiveddate{XXXX}
\Reviseddate{XXXX}
\Accepteddate{XXXX}
\Dateposted{XXXX}



\title[Equivalence of the effective Hamiltonian approach and the Siegert boundary condition]{Equivalence of the effective Hamiltonian approach and the Siegert boundary condition for resonant states}


\author[N. Hatano]{Naomichi Hatano
  \footnote{Corresponding author\quad E-mail:~\textsf{hatano@iis.u-tokyo.ac.jp},
            Phone: +81\,3\,5452\,6154,
            Fax: +81\,3\,5452\,6155}}
\address[]{Institute of Industrial Science, University of Tokyo, Komaba 4-6-1, Meguro, Tokyo 153-8505, Japan}
\begin{abstract}
Two theoretical methods of finding resonant states in open quantum systems, namely the approach of the Siegert boundary condition and the Feshbach formalism, are reviewed and shown to be algebraically equivalent to each other for a simple model of the T-type quantum dot.
It is stressed that the seemingly Hermitian Hamiltonian of an open quantum system is implicitly non-Hermitian outside the Hilbert space.
The two theoretical approaches extract an explicitly non-Hermitian effective Hamiltonian in a contracted space out of the seemingly Hermitian (but implicitly non-Hermitian) full Hamiltonian in the infinite-dimensional state space of an open quantum system. 
\end{abstract}
\maketitle                   





\section{Introduction}

Resonance phenomena are prevalent in physics; see \textit{e.g.}\ Ref.~\cite{Prigogine96}.
Quantum resonance, in particular, has been a central issue of quantum mechanics from the early days of its development~\cite{Gamow28,Eckart30,Bethe36,Siegert39,Hulthen42a,Hulthen42b,Jost51,Vogt54,Corinaldesi56,Peierls59,Nussenzveig59,Landau77,Brandas90,Kukulin89}.
It is ever more important these days because we can closely observe and even manipulate quantum-mechanical systems such as mesoscopic systems, molecules and nuclides~\cite{Kobayashi02,Kobayashi03,Kobayashi04,Sato05,Brisker08,Morita04}:
for example, the resonant conduction in experiments of mesoscopic systems yields a variety of functions of nano-devices;
producing unstable nuclides experimentally is nothing but looking for resonant states.
It may be said, however, that such experimental studies on quantum resonance far surpass theoretical ones.
Further development of theoretical and numerical methods of analyzing quantum resonance is awaited.

In the present note, we review two theoretical methods of finding resonant states in open quantum systems, namely the approach of the Siegert boundary condition~\cite{Siegert39,Landau77,Hatano08,Nakamura07,Hatano09,Garmon09,Klaiman11} and the Feshbach formalism~\cite{Feshbach58,Feshbach62,Sadreev03,Rotter09}.
One of the purposes of the note is to show the algebraic equivalence of the two approaches.

Another point of the note is to stress that the full Hamiltonian of an open quantum system is Hermitian only in the Hilbert space, which is spanned by the bound states and the scattering continuum states.
It is generally non-Hermitian outside the Hilbert space, where the resonant states reside~\cite{Hatano08,Aguilar71,Balslev71,Simon72}.
That is why the resonant states can have complex eigenvalues of the seemingly Hermitian Hamiltonian.
The two numerical methods named above produce an explicitly non-Hermitian effective Hamiltonian in a contracted finite-dimensional state space out of the implicitly non-Hermitian full Hamiltonian in the infinite-dimensional state space.

The note is organized as follows.
First in the next section~\ref{sec2}, we introduce the simple model that we are going to use in the present note.
We review the approach of the Siegert boundary condition in Sec.~\ref{sec5}.
We then go on to the Feshbach formalism in Sec.~\ref{sec3}.
We finally present in Sec.~\ref{sec7} the statement that the full Hamiltonian is Hermitian in the Hilbert space, whereas non-Hermitian outside the Hilbert space.

\section{Model}
\label{sec2}

\begin{figure}
\centering
\includegraphics[width=0.5\textwidth]{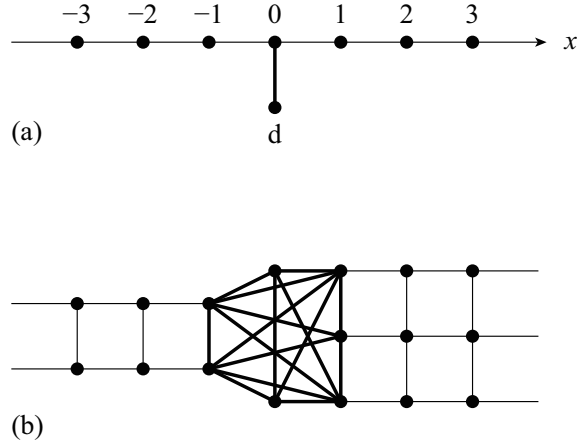}
\caption{(a) The tight-binding model of a T-type dot.
The sites on the infinite lead are labeled by integers $0,\pm1,\pm2,\ldots$, whereas the site on the dot is labeled by ``$\rmd$."
(b) A general open quantum system to which we can generalize the present argument.}
\label{fig1}
\end{figure}

Let us consider hereafter the simple model in Fig.~\ref{fig1}(a) for concreteness.
It is sometimes called a T-type dot or a side-coupled dot~\cite{Porod92,Porod93,Shao94,Kim01,Kikoin01,Kang01,Affleck01,Simon01,Affleck07}.
We can obviously extend the following arguments to more general models such as the one in Fig.~\ref{fig1}(b)~\cite{Datta95,Sasada05,Sasada11}.

We will leave out any inter-particle interactions;
see \textit{e.g.}\ Refs.~\cite{Nishino07,Imamura09,Nishino11} for quantum-dot systems Coulombic interactions.
The present problem is strictly a one-body problem.
The whole state space is hence spanned by the state $|\rmd\rangle$ in which the particle resides on the dot site $\rmd$ and the state $|x\rangle$ in which the particle resides on the site $x$ of the lead, where $x$ runs from $-\infty$ to $+\infty$.
The resolution of unity is given by
\begin{align}
|\rmd\rangle\langle\rmd|+\sum_{x=-\infty}^\infty|x\rangle\langle x|=1.
\end{align}

The Hamiltonian of the T-type dot is expressed in the form
\begin{align}\label{eq10}
H&=-t\sum_{x=-\infty}^\infty
\left(|x+1\rangle\langle x|+|x\rangle\langle x+1|\right)
-t_1\left(|\rmd\rangle\langle 0|+|0\rangle\langle\rmd|\right)
+\varepsilon_\rmd|\rmd\rangle\langle\rmd|,
\end{align}
where $t$ is the transfer amplitude between neighboring sites on the lead, $t_1$ is the one between the site $x=0$ and the site $\rmd$, and $\varepsilon_\rmd$ is the potential at the dot site $\rmd$.

The issue here is to find resonance poles of the S matrix of this open quantum system.
In the early days of quantum mechanics, it was noticed that we can find the resonance pole as an eigenstate of the time-independent Schr\"{o}dinger equation under the Siegert boundary condition~\cite{Siegert39}.
Later on, Feshbach~\cite{Feshbach58,Feshbach62} as well as other researchers~\cite{Petrosky91} established the Feshbach formalism, which algebraically produces resonance poles.
In the present paper we will review the two methods respectively in Secs.~\ref{sec5} and~\ref{sec3}, in the course of which we will show their equivalence.

\section{Siegert boundary condition}
\label{sec5}

We first review the approach of the Siegert boundary condition, the condition that the resonant state has only out-going waves away from the scatterer.
Let us first explain how this condition comes out.

In solving the scattering problem in one dimension, we typically assume the incoming wave, the reflection wave and the transmission wave in the form
\begin{align}\label{eq30-1}
\Phi(x)=
\begin{cases}
Ae^{ikx}+Be^{-ikx} & \quad\mbox{on the left of the scatterer,}\\
Ce^{ikx} &\quad\mbox{on the right of the scatterer.}
\end{cases}
\end{align}
The reflection and transmission amplitudes, which are elements of the S matrix, are given by $B/A$ and $C/A$, respectively.
A textbook definition of the resonant state as a pole of the S matrix then implies that we should look for the zeros of the amplitude $A$ of the incoming wave.
We can do so by making the amplitude $A$ zero from the very beginning:
\begin{align}\label{eq40-1}
\Phi(x)=
\begin{cases}
Be^{-ikx} & \quad\mbox{on the left of the scatterer,}\\
Ce^{ikx} &\quad\mbox{on the right of the scatterer.}
\end{cases}
\end{align}
This is the Siegert boundary condition that the resonant state has only out-going waves away from the scatterer~\cite{Siegert39,Landau77,Hatano08,Klaiman11}.
In fact, the Siegert boundary condition not only supports the resonant states but all other possible discrete states including the bound states and the anti-resonant states, that is, all poles of the S matrix.
A bound state follows from a pure imaginary value of $k$.
A resonant state comes out for $\mathop{\mathrm{Re}}k>0$, whereas 
an anti-resonant state comes out for $\mathop{\mathrm{Re}}k<0$.

Let us then solve the Schr\"{o}dinger equation
\begin{align}\label{eq50-1}
H|\Phi\rangle=E|\Phi\rangle
\end{align}
under the Siegert boundary condition~\eqref{eq40-1}, specifically for the T-type dot~\eqref{eq10}, under the condition
\begin{align}\label{eq220}
\langle x|\Phi\rangle=
\begin{cases}
Be^{-ikx}&\quad\mbox{for }x\leq-1,
\\
Ce^{ikx}&\quad\mbox{for }x\geq1.
\end{cases}
\end{align}
First, the Schr\"{o}dinger equation~\eqref{eq50-1} for the sites $x\leq -2$ and $x\geq 2$ is written in the form
\begin{align}\label{eq150}
-t\left(\langle x-1|\Phi\rangle+\langle x+1|\Phi\rangle\right)
=E\langle x|\Phi\rangle,
\end{align}
which along with the condition~\eqref{eq220} yields the dispersion relation
\begin{align}\label{eq130}
E=-t\left(e^{ik}+e^{-ik}\right)=-2t\cos k
\end{align}
of the lead.
Next, the Schr\"{o}dinger equation~\eqref{eq50-1} for the sites $x=\pm 1$ reads
\begin{align}\label{eq90-1}
-t\left(\langle-2|\Phi\rangle+\langle0|\Phi\rangle\right)&=E\langle-1|\Phi\rangle,
\\\label{eq100-1}
-t\left(\langle0|\Phi\rangle+\langle2|\Phi\rangle\right)&=E\langle1|\Phi\rangle.
\end{align}
By using the condition~\eqref{eq220} with the dispersion relation~\eqref{eq130}, we simply obtain
the continuity condition
\begin{align}
B=C=\langle0|\Phi\rangle.
\end{align}

Finally for the remaining sites $x=0$ and $\rmd$, the Schr\"{o}dinger equation~\eqref{eq50-1} reads
\begin{align}\label{eq120-1}
-t\left(\langle-1|\Phi\rangle+\langle1|\Phi\rangle\right)-t_1\langle\rmd|\Phi\rangle
&=E\langle0|\Phi\rangle,
\\\label{eq130-1}
-t_1\langle0|\Phi\rangle+\varepsilon_\rmd\langle\rmd|\Phi\rangle
&=E\langle\rmd|\Phi\rangle.
\end{align}
The point here is the fact that the Siegert condition~\eqref{eq220} yields~\cite{Sasada08}
\begin{align}\label{eq140-2}
\langle\pm1|\Phi\rangle=e^{ik}\langle0|\Phi\rangle.
\end{align}
We thereby arrive at a \textit{closed} set of equations of the form
\begin{align}
-2te^{ik}\langle0|\Phi\rangle-t_1\langle\rmd|\Phi\rangle
&=E\langle0|\Phi\rangle,
\\
-t_1\langle0|\Phi\rangle+\varepsilon_\rmd\langle\rmd|\Phi\rangle
&=E\langle\rmd|\Phi\rangle.
\end{align}
These two equations are cast into the matrix form
\begin{align}\label{eq320}
\begin{pmatrix}
-2te^{ik}&-t_1\\
-t_1&\varepsilon_\rmd
\end{pmatrix}
\begin{pmatrix}
\langle0|\Phi\rangle\\
\langle\rmd|\Phi\rangle
\end{pmatrix}
=E
\begin{pmatrix}
\langle0|\Phi\rangle\\
\langle\rmd|\Phi\rangle
\end{pmatrix}.
\end{align}
We will indeed show in Sec.~\ref{sec3} that the two-by-two matrix on the left-hand side is the non-Hermitian effective Hamiltonian given below by the Feshbach formalism:
\begin{align}\label{eq180-2}
\Heff=\begin{pmatrix}
-2te^{ik}&-t_1\\
-t_1&\varepsilon_\rmd
\end{pmatrix}.
\end{align}
Equation~\eqref{eq320} is then cast into the form of the eigenproblem for the effective Hamiltonian:
\begin{align}\label{eq190-1}
\Heff(P|\Phi\rangle)=E(P|\Phi\rangle),
\end{align}
where $P$ is a projection operator defined below, but for the present purpose it is enough to say
\begin{align}
P|\Phi\rangle=
\begin{pmatrix}
\langle0|\Phi\rangle\\
\langle\rmd|\Phi\rangle
\end{pmatrix}.
\end{align}
Once we solve the eigenproblem~\eqref{eq190-1}, the wave functions away from the scatterer is given by the Siegert condition~\eqref{eq220} as
\begin{align}\label{eq330}
\langle x|\Phi\rangle=e^{ik|x|}\langle 0|\Phi\rangle.
\end{align}

The fact that the effective Hamiltonian of the Feshbach formalism emerges when we solve the scattering problem of the full Hamiltonian is not restricted to the Siegert boundary condition, but is prevalent in other boundary conditions.
See Appendix~\ref{appA} below.


To summarize the approach of the Siegert boundary condition, we begin with solving the original Schr\"{o}dinger equation~\eqref{eq50-1} for the \textit{full} Hamiltonian
under the Siegert boundary condition~\eqref{eq220}.
We then end up with the Schr\"{o}dinger equation~\eqref{eq190-1} for the effective Hamiltonian, augmented by Eq.~\eqref{eq330}.
These are exactly the same as the equations that will be derived from the Feshbach formalism in Sec.~\ref{sec3}.

We thereby obtain discrete eigenvalues including resonant ones by finding the roots of
\begin{align}\label{eq260}
\det (E-\Heff)=0
\end{align}
in the complex $E$ plane.
Note here that the number of the roots can be more than two although the effective Hamiltonian~\eqref{eq210-1} is a two-by-two matrix, because the Hamiltonian itself depends on the energy $E$ through the dispersion relation~\eqref{eq130}.
Using the variable $z=e^{ik}$ with the mapping
\begin{align}
E=-t(z+z^{-1}),
\end{align}
we can indeed transform the secular equation~\eqref{eq260} into a fourth-order polynomial equation with respect to $z$,
\begin{align}
t^2z^4+t\varepsilon_\rmd z^3 +{t_1}^2z^2-t\varepsilon_\rmd z-t^2=0,
\end{align}
in the particular case of Eq.~\eqref{eq320}. 
Its four solutions contain two bound states, one resonant state and one anti-resonant state in one parameter region, while two bound states and two anti-bound states in the other regions~\cite{Hatano08}.

We will show in Sec.~\ref{sec7} why the effective Hamiltonian~\eqref{eq180-2} can be non-Hermitian and why complex eigenvalues can emerge in spite of the fact that they come out of the seemingly Hermitian full Hamiltonian~\eqref{eq10}.
We will argue that the full Hamiltonian of an open quantum system is in fact non-Hermitian outside the Hilbert space and that the effective Hamiltonian~\eqref{eq180-2} manifests the non-Hermiticity.

\section{The Feshbach formalism of the effective Hamiltonian}
\label{sec3}

Let us now review the Feshbach formalism for the T-type dot~\eqref{eq10}~\cite{Sadreev03,Rotter09}, which will produce the equations exactly the same as Eqs.~\eqref{eq190-1} and~\eqref{eq330}.

\begin{figure}
\centering
\includegraphics[width=0.5\textwidth]{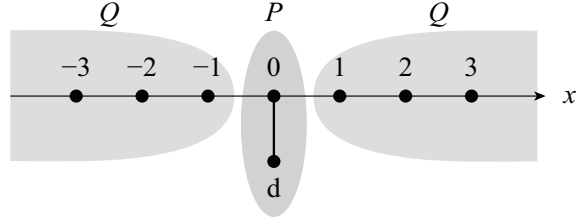}
\caption{The subspaces projected by the operators $P$ and $Q$.}
\label{fig2}
\end{figure}

We introduce the projection operators $P$ and $Q$ as follows:
\begin{align}
\Ps&=|\rmd\rangle\langle\rmd|+|0\rangle\langle0|,
\\
\Qs&=1-\Ps
=\sum_{x=1}^\infty |x\rangle\langle x|+\sum_{x=-1}^{-\infty} |x\rangle\langle x|.
\end{align}
This divides the T-type dot in Fig.~\ref{fig1}(a) into the subspaces indicated in Fig.~\ref{fig2}.

We would like to solve the Schr\"{o}dinger equation~\eqref{eq50-1} for the full Hamiltonian.
The Feshbach formalism tells us that instead of solving Eq.~\eqref{eq50-1} in the state space in the infinite dimensions, we can solve
\begin{align}\label{eq100}
\Heff\Ps|\Phi\rangle=E\Ps|\Phi\rangle
\end{align}
in the $\Ps$ subspace in two dimensions.
We are going to show in generic cases that the Feshbach formalism gives the effective Hamiltonian $\Heff$ in Eq.~\eqref{eq100} in the form
\begin{align}\label{eq110}
\Heff=\Ps H\Ps+\Ps H\Qs\frac{1}{E-\Qs H\Qs}\Qs H\Ps.
\end{align}
Note that we can express the effective Hamiltonian as a matrix in the $\Ps$ subspace.

Here is how we can obtain the reduced Schr\"{o}dinger equation~\eqref{eq100} with the effective Hamiltonian~\eqref{eq110}.
We can split the original Schr\"{o}dinger equation~\eqref{eq50-1} into the $\Ps$ and $\Qs$ subspaces as
\begin{align}
\Ps H(\Ps+\Qs)|\Phi\rangle&=E\Ps|\Phi\rangle,
\\
\Qs H(\Ps+\Qs)|\Phi\rangle&=E\Qs|\Phi\rangle,
\end{align}
which are cast into the forms
\begin{align}\label{eq140-1}
\Ps H\Ps(\Ps|\Phi\rangle)+\Ps H\Qs(\Qs|\Phi\rangle)&=E(\Ps|\Phi\rangle),
\\\label{eq150-1}
\Qs H\Ps(\Ps|\Phi\rangle)+\Qs H\Qs(\Qs|\Phi\rangle)&=E(\Qs|\Phi\rangle).
\end{align}
We first solve the second equation~\eqref{eq150-1} with respect to the state $\Qs|\Phi\rangle$ as
\begin{align}\label{eq160-1}
\Qs|\Phi\rangle=\frac{1}{E-\Qs H\Qs}\Qs H\Ps(\Ps|\Phi\rangle)
\end{align}
and substitute Eq.~\eqref{eq160-1} into the first equation~\eqref{eq140-1} to obtain
\begin{align}
\Ps H\Ps(\Ps|\Phi\rangle)+\Ps H\Qs\frac{1}{E-\Qs H\Qs}\Qs H\Ps(\Ps|\Phi\rangle)=E(\Ps|\Phi\rangle).
\end{align}
We can cast this into the form~\eqref{eq100} with the effective Hamiltonian~\eqref{eq110}.

Specifically for the T-type dot~\eqref{eq10}, each term in the effective Hamiltonian~\eqref{eq110} is expressed as follows:
\begin{align}\label{eq40}
\Ps H\Ps&=-t_1\left(|\rmd\rangle\langle 0|+|0\rangle\langle\rmd|\right)
+\varepsilon_\rmd|\rmd\rangle\langle\rmd|,
\\
\Qs H\Ps&=-t|1\rangle\langle0|-t|-1\rangle\langle0|.
\\\label{eq80}
\Qs H\Qs&=-t\sum_{x=1}^\infty
\left(|x+1\rangle\langle x|+|x\rangle\langle x+1|\right)
-t\sum_{x=-1}^{-\infty}
\left(|x-1\rangle\langle x|+|x\rangle\langle x-1|\right)
\\
\Ps H\Qs&=-t|0\rangle\langle1|-t|0\rangle\langle-1|,
\end{align}
These expressions enable us to understand why the finite-dimensional effective Hamiltonian~\eqref{eq110} can represent the dynamics of the particle in the infinite-dimensional space.
The first term $\Ps H\Ps$ of the effective Hamiltonian indeed describes the dynamics only in the $\Ps$ subspace of the sites d and 0.
Its second term, on the other hand, describes the following dynamics:
the particle first hops from the $\Ps$ subspace to the $\Qs$ subspace owing to the partial Hamiltonian $\Qs H\Ps$;
the particle moves around in the infinite-dimensional $\Qs$ subspace under the Green's function of the the partial Hamiltonian $\Qs H\Qs$;
the particle then hops from the $\Qs$ subspace back to the $\Ps$ subspace owing to the partial Hamiltonian $\Ps H\Qs$.
The fact that the system is open is thus encoded in the second term of the effective Hamiltonian~\eqref{eq110}.
This term indeed turns out to be non-Hermitian.

We can calculate the Green's function $(E-QHQ)^{-1}$ explicitly for the partial Hamiltonian~\eqref{eq80}.
We do not give the details of the calculation~\cite{Sasada11} but present the following result:
\begin{align}\label{eq180-1}
\left.\begin{array}{ll}
\mbox{for } x\geq 1\quad &\displaystyle \langle x|\frac{1}{E-QHQ}|1\rangle
\\
&\\
\mbox{for } x\leq -1\quad &\displaystyle\langle x|\frac{1}{E-QHQ}|-1\rangle
\end{array}\right\}
=-\frac{1}{t}e^{ik|x|}.
\end{align}
Here we express the right-hand side in terms of the wave number $k$ on the lead instead of the energy $E$;
we indeed obtain the dispersion relation $E=-2t\cos k$ during the calculation of Eq.~\eqref{eq180-1} as the eigenvalues of the partial Hamiltonian $QHQ$.
The wave number $k$, or more precisely the crystal wave number, is restricted to the first Brillouin zone $-\pi\leq k\leq\pi$ for the energy band $-2t\leq E\leq 2t$, where the point $k=-\pi$ is equivalent to the point $k=\pi$ because of the $2\pi$ periodicity of the Fourier space.
The variables $k$ and $E$ are related through the dispersion relation of the lead, Eq.~\eqref{eq130}.
The mapping from the energy $E$ to the wave number $k$ is actually a one-to-two correspondence.
We obtain the retarded Green's function out of Eq.~\eqref{eq180-1} by choosing the branch $0<k< \pi$ and the advanced one by choosing the branch $-\pi<k<0$.

Using the Green's function~\eqref{eq180-1}, we have
\begin{align}\label{eq380-1}
PHQ\frac{1}{E-QHQ}QHP&=
(-t)^2|0\rangle\langle1|\frac{1}{E-QHQ}|1\rangle\langle0|
\nonumber\\
&+
(-t)^2|0\rangle\langle-1|\frac{1}{E-QHQ}|-1\rangle\langle0|
\nonumber\\
&=-2te^{ik}|0\rangle\langle0|.
\end{align}
Therefore, we can express the effective Hamiltonian~\eqref{eq110} in the following form of the two-by-two matrix spanned by the states $|0\rangle$ and $|\rmd\rangle$~\cite{Sadreev03}:
\begin{align}\label{eq210-1}
\Heff=\begin{pmatrix}
-2te^{ik} & -t_1 \\
-t_1 & \varepsilon_\rmd
\end{pmatrix}.
\end{align}



To summarize the effective-Hamiltonian approach, we solve the Schr\"{o}dinger equation of the non-Hermitian effective Hamiltonian, Eq.~\eqref{eq100}, which is equivalent to Eq.~\eqref{eq190-1}.
%
We can obtain the corresponding eigenfunctions as follows.
The $\Ps$ part of the resonant wave function naturally follows from the eigenproblem~\eqref{eq100}.
We can obtain the part $Q|\Phi\rangle$ from $P|\Phi\rangle$ as
\begin{align}
Q|\Phi\rangle=\frac{1}{E-QHQ}QHP|\Phi\rangle,
\end{align}
or more specifically,
\begin{align}\label{eq210}
\langle x|\Phi\rangle&=
\begin{cases}
\displaystyle
\langle x|\frac{1}{E-QHQ}|1\rangle\langle0|\Phi\rangle & \quad\mbox{for } x\geq 1,
\\
&\\
\displaystyle
\langle x|\frac{1}{E-QHQ}|-1\rangle\langle0|\Phi\rangle & \quad\mbox{for } x\leq -1
\end{cases}
\nonumber\\
\nonumber\\
&=e^{ik|x|}\langle 0|\Phi\rangle,
\end{align}
which is the same as Eq.~\eqref{eq330}.
Therefore, the approach of the Siegert boundary condition and the Feshbach formalism are algebraically equivalent and produce the same eigenstates.

\section{Origin of the non-Hermiticity of the effective Hamiltonian}
\label{sec7}

\begin{figure}
\centering
\includegraphics[width=0.5\textwidth]{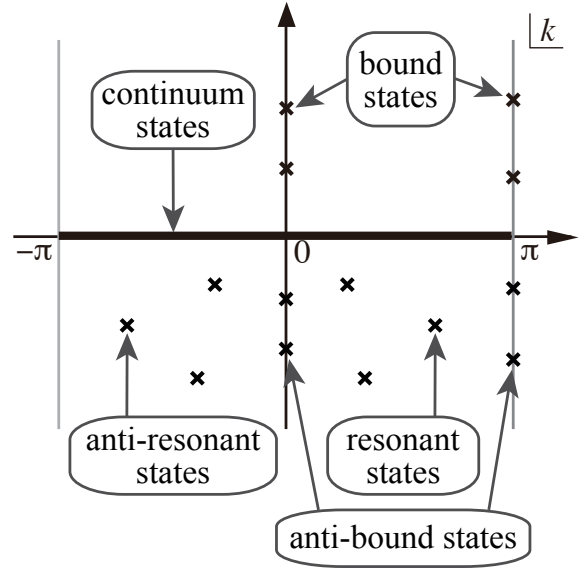}
\caption{The distribution of eigenstates of the Hamiltonian~\eqref{eq10} in the complex $k$ plane.}
\label{fig3}
\end{figure}

Let us argue by examining the functional space of the eigenstates, that the full Hamiltonian $H$ is \textit{non-Hermitian outside the Hilbert space}.
The eigenstates obtained either with the Siegert boundary condition or in the Feshbach formalism are distributed in the complex $k$ plane as shown in Fig.~\ref{fig3}.
Specifically for the T-type dot~\eqref{eq10} with $\varepsilon_\rmd=0$ for example, the solutions are
\begin{align}
\label{eqC110}
&
k^\mathrm{bl}=0+i\log p,
\qquad
E^\mathrm{bl}=-(p+q)t,
\\\label{eqC120}
&
k^\mathrm{bh}=\pi+i\log p,
\qquad
E^\mathrm{bh}=(p+q)t,
\\\label{eqC180}
&
k^\mathrm{res}=+\frac{\pi}{2}-i\log p,
\qquad
E^\mathrm{res}=-i(p-q)t,
\\\label{eqC190}
&
k^\mathrm{ar}=-\frac{\pi}{2}-i\log p,
\qquad
E^\mathrm{ar}=+i(p-q)t,
\end{align}
where 
\begin{align}
p&=\sqrt{\frac{{t_1}^2+\sqrt{4t^4+{t_1}^4}}{2t^2}},
\\
q&=\sqrt{\frac{-{t_1}^2+\sqrt{4t^4+{t_1}^4}}{2t^2}}=\frac{1}{p}
\end{align}
are dimensionless real numbers.
The superscripts `bl,' `bh,' `res' and `ar' in Eqs.~\eqref{eqC110}--\eqref{eqC190} respectively denote the bound state with $\mathop{\mathrm{Re}}k=0$, the bound state with $\mathop{\mathrm{Re}}k=\pi$, the resonant state and the anti-resonant state.
The distribution for the tight-binding model~\eqref{eq10} is different from the standard distribution in two points~\cite{Sasada11}:
first, the whole $k$ plane is limited to $-\pi<\mathop{\mathrm{Re}}k\leq\pi$ because of the lattice periodicity;
second, as a consequence of the first point, bound and anti-bound states can exist, respectively, on the positive and negative imaginary parts of the line $\mathop{\mathrm{Re}}k=\pi$, which correspond to the energy eigenvalues above the upper threshold $E=+2t$~\cite{Garmon09}.
Other properties are the same as the standard distribution.

The upper and lower half $k$ planes, respectively, correspond to the first and second Riemann sheets of the complex $E$ plane.
First, the bound states $\{|\Phi_n^\textrm{b}|n=1,2,3,\ldots\rangle\}$ are distributed in the upper half $k$ plane.
Their eigenfunctions are localized around the scatterer.
When we substitute $k=i\kappa$ with $\kappa>0$, the wave function~\eqref{eq330} is of the form
\begin{align}
\langle x|\Phi\rangle=e^{-\kappa|x|}\langle 0|\Phi\rangle,
\end{align}
and is $L^2$-normalizable.
Next, the continuum scattering states $\{|\Phi_k\rangle|-\pi<k\leq\pi\}$ are on the real $k$ axis.
Their eigenfunctions are plane waves of the form
\begin{align}
\langle x|\Phi\rangle=Ae^{ikx}+B^{-ikx}
\end{align}
with real $k$, and hence box-normalizable.
Indeed, it was proved~\cite{Newton61} that the above two types of the normalizable eigenfunctions form a complete set
\begin{align}
\sum_{n}
|\Phi_n^\textrm{b}\rangle\langle\Phi_n^\textrm{b}|
+\int_{-\pi}^\pi\frac{dk}{2\pi}|\Phi_k\rangle\langle\Phi_k|=1.
\end{align}
We therefore refer to the functional space spanned by these normalizable eigenstates as the Hilbert space.

On the other hand, the resonant and anti-resonant states as well as the anti-bound states are all distributed in the lower half $k$ plane, or the second Riemann sheet.
Their wave functions are not normalizable in the usual sense, although there is a method of regularization called complex scaling~\cite{Aguilar71,Balslev71,Simon72,Homma97,Masui99,Moiseyev08}.
When we substitute $k=k_\textrm{r}+i\kappa$ with $\kappa<0$, the wave function~\eqref{eq330} is of the form
\begin{align}
\langle x|\Phi\rangle\propto e^{-\kappa|x|}=e^{+|\kappa||x|},
\end{align}
which is spatially divergent~\cite{leCouteur60,Zeldovich60,Hokkyo65,Romo68,Berggren68,Berggren70,Gyarmati71,Romo80,Berggren82,Berggren96,Madrid05,Hatano08,Hatano09}.
The point here is that the wave functions of the resonant, anti-resonant and anti-bound states are not normalizable and exist outside the Hilbert space.
(We argued~\cite{Hatano08,Hatano09} nonetheless that this spatial divergence combined with the temporal decay (due to the imaginary part of the eigenenergy) is followed by the probabilistic interpretation of the resonant wave function.)

To summarize, the full Hamiltonian~\eqref{eq10} is Hermitian in the Hilbert space, whereas non-Hermitian outside the Hilbert space.
Indeed, all eigenstates in the Hilbert space (the bound states and the scattering continuum) have real eigenvalues, whereas the eigenvalues of the states outside the Hilbert space (the resonant and anti-resonant states) are generally complex (with the exception of the anti-bound states).

From this viewpoint, the effective Hamiltonian explicitly reveals the implicit non-Hermiticity of the full Hamiltonian.
In the approach of the Siegert boundary condition, the non-Hermiticity of the effective Hamiltonian arises in Eq.~\eqref{eq140-2}, which comes from the Siegert boundary condition~\eqref{eq40-1}.
In the Feshbach formalism, the non-Hermiticity is due to the term~\eqref{eq380-1}.
In both cases, we can say that the explicit non-Hermiticity of the effective Hamiltonian originates from the openness of the system, which is also the origin of the implicit non-Hermiticity of the full Hamiltonian.

\section{Summary}

In the present note, we showed the equivalence of the approach of the Siegert boundary condition and the Feshbach formalism.
We can also show the equivalence among the standard scattering boundary condition~\eqref{eq30-1}, the Feshbach formalism, and the Lippmann-Schwinger equation, which we omitted here.

An important message here is that the Hamiltonian of an open quantum system is, although it is Hermitian in the Hilbert space, generally non-Hermitian outside the Hilbert space.
The effective Hamiltonian manifests the non-Hermiticity in a contracted state space.

\appendix
\section{Other incidents that the effective Hamiltonian appears in the scattering problem of the full Hamiltonian}
\label{appA}

The fact that the effective Hamiltonian of the Feshbach formalism emerges when we solve the scattering problem of the full Hamiltonian is not restricted to the Siegert boundary condition~\eqref{eq40-1}, but is prevalent in other boundary conditions.

\subsection{The standard scattering problem}
First, let us consider the standard scattering solution
\begin{align}\label{eqA10}
\langle x|\Phi\rangle=
\begin{cases}
Ae^{ikx}+Be^{-ikx}&\quad\mbox{for }x\leq-1,
\\
Ce^{ikx}&\quad\mbox{for }x\geq1.
\end{cases}
\end{align}
We then follow the steps outlined in Sec.~\ref{sec5}.
First for $x\leq-2$ and $x\geq2$, the Schr\"{o}dinger equation~\eqref{eq50-1} reads Eq.~\eqref{eq150}, which is only followed by the dispersion relation~\eqref{eq130}.
Next for $x=\pm 1$, the Schr\"{o}dinger equation~\eqref{eq50-1} results in Eqs.~\eqref{eq90-1} and~\eqref{eq100-1}.
By using the scattering condition~\eqref{eqA10} with the dispersion relation~\eqref{eq130}, we simply obtain the continuity condition
\begin{align}
A+B=C=\langle0|\Phi\rangle.
\end{align}
Finally for the site $0$ and the dot site $\rmd$, the Schr\"{o}dinger equation~\eqref{eq50-1} gives Eqs.~\eqref{eq120-1} and~\eqref{eq130-1}.
Here the standard scattering condition~\eqref{eqA10} gives
\begin{align}
\langle-1|\Phi\rangle&=Ae^{-ik}+\left(\langle0|\Phi\rangle-A\right)e^{ik}
\nonumber\\
&=-2iA\sin k+e^{ik}\langle0|\Phi\rangle,
\\
\langle1|\Phi\rangle&=Ce^{ik}=e^{ik}\langle0|\Phi\rangle.
\end{align}
Therefore we have
\begin{align}
-2te^{ik}\langle0|\Phi\rangle-t_1\langle\rmd|\Phi\rangle
&=E\langle0|\Phi\rangle-2itA\sin k,
\\
-t_1\langle0|\Phi\rangle+\varepsilon_\rmd\langle\rmd|\Phi\rangle
&=E\langle\rmd|\Phi\rangle.
\end{align}

These two equations are cast into the matrix form
\begin{align}
\left(
E-\Heff
\right)
\begin{pmatrix}
\langle0|\Phi\rangle\\
\langle\rmd|\Phi\rangle
\end{pmatrix}
=A
\begin{pmatrix}
2it\sin k\\
0
\end{pmatrix},
\end{align}
or
\begin{align}\label{eqA200}
\begin{pmatrix}
\langle0|\Phi\rangle\\
\langle\rmd|\Phi\rangle
\end{pmatrix}
=\frac{A}{E-\Heff}
\begin{pmatrix}
2it\sin k\\
0
\end{pmatrix}.
\end{align}
Note that the effective Hamiltonian $\Heff$ appeared here not from the Feshbach formalism, but by solving the Schr\"{o}dinger equation of the full Hamiltonian.
Incidentally, the quantity $2t\sin k$ on the right-hand side is the group velocity according to the dispersion relation~\eqref{eq130}:
\begin{align}
v_\textrm{g}\equiv\frac{dE}{dk}=2t\sin k.
\end{align}

\subsection{Green's function}

The Green's function with the source at the site $0$ is the solution of 
\begin{align}\label{eqA110}
\left(E-H\right)G_{x0}=\delta_{x0}.
\end{align}
The retarded Green's function is given under the boundary condition
\begin{align}
G_{x0}=\begin{cases}
e^{ikx}&\quad\mbox{for }x\geq0,
\\
e^{-ikx}&\quad\mbox{for }x\leq0.
\end{cases}
\end{align}
Similar steps as above reduce Eq.~\eqref{eqA110} to
\begin{align}
\left(
E-\Heff
\right)
\begin{pmatrix}
G_{00}\\
G_{\rmd0}
\end{pmatrix}
=
\begin{pmatrix}
1\\
0
\end{pmatrix},
\end{align}
or
\begin{align}
\begin{pmatrix}
G_{00}\\
G_{\rmd0}
\end{pmatrix}
=
\frac{1}{E-\Heff}\begin{pmatrix}
1\\
0
\end{pmatrix}.
\end{align}
Therefore, we can describe the scattering solution~\eqref{eqA200} in terms of the Green's function as
\begin{align}
\begin{pmatrix}
\langle0|\Phi\rangle\\
\langle\rmd|\Phi\rangle
\end{pmatrix}
=iAv_\textrm{g}\begin{pmatrix}
G_{00}\\
G_{\rmd0}
\end{pmatrix}.
\end{align}

\begin{acknowledgement}
The author is very much in debt to Dr.~Ingrid~Rotter for valuable discussions.
The present work is supported by the Grant-in-Aid for Scientific Research No.~22 340 110 from MEXT, Japan.
\end{acknowledgement}


\providecommand{\WileyBibTextsc}{}
\let\textsc\WileyBibTextsc
\providecommand{\othercit}{}
\providecommand{\jr}[1]{#1}
\providecommand{\etal}{~et~al.}

\end{document}